\documentclass[a4paper,fleqn]{cas-dc}

\usepackage{float}
\usepackage{tcolorbox}
\usepackage{placeins}
\usepackage[inline]{enumitem} % this is only used for figure 1 caption.
\tcbuselibrary{listings,breakable,skins,raster}

\definecolor{chat-colback}{HTML}{ffffff}%{ffecd1}
\definecolor{chat-colframe}{HTML}{003049}%{001524}%{14213d}
\definecolor{chat-colbacklower}{HTML}{eae2b7}
\definecolor{chatmulti-colback}{HTML}{f5f5f5}
\definecolor{chatgpt-colframe}{HTML}{d62828}
\definecolor{promptbox-colframe}{HTML}{fcbf49}
\tcbsetforeverylayer{colback=chat-colback, colframe=chat-colframe,skin=bicolor, colbacklower=chat-colbacklower!60!white}

\usepackage[numbers]{natbib}
\newcommand{\fakeref}[1]{\textcolor{chat-colframe}{#1}}

\ExplSyntaxOn
\cs_gset:Npn \__first_footerline:
  { \group_begin: \small \sffamily \__short_authors: \group_end: }
\ExplSyntaxOff

\begin{document}

% \newtcolorbox{ExampleChat}[1]{colframe=red!75!black,colback=red!5!white,#1}

\let\WriteBookmarks\relax
\def\floatpagepagefraction{1}
\def\textpagefraction{.001}

% Short title
\shorttitle{AI Language Models and Marine Policy Equity}

% Short author
\shortauthors{Ziegler, Lothain, O'Neill, Anderson, Ota}

% Main title of the paper 
\title [mode = title]{AI Language Models Could Both Help and Harm Equity in Marine Policymaking: The Case Study of the BBNJ Question-Answering Bot}   

\let\printorcid\relax % Remove ORCID footnote

% First author
%
% Options: Use if required
% eg: \author[1,3]{Author Name}[type=editor,
%       style=chinese,
%       auid=000,
%       bioid=1,
%       prefix=Sir,
%       orcid=0000-0000-0000-0000,
%       facebook=<facebook id>,
%       twitter=<twitter id>,
%       linkedin=<linkedin id>,
%       gplus=<gplus id>]
\author[2,1]{Matt Ziegler}\cormark[1]\ead{mattzig@cs.washington.edu}
\author[3,1]{Sarah Lothian}
\author[2,1]{Brian O'Neill}
\author[2,1]{Richard Anderson}
\author[1,2]{Yoshitaka Ota}

\affiliation[1]{organization={Ocean Nexus Center},city={Seattle},country={USA}}
\affiliation[2]{organization={University of Washington}, city={Seattle}, country={USA}}
\affiliation[3]{organization={University of Wollongong}, city={Wollongong}, country={Australia}}

\cortext[cor1]{Corresponding author}

% Here goes the abstract
\begin{abstract}
AI Large Language Models (LLMs) like ChatGPT have exploded in popularity and are set to reshape some aspects of policymaking processes. Despite their poorly-understood risks, policy practitioners are already using ChatGPT for help with a variety of tasks: from drafting statements, submissions, and presentations, to conducting background research. We are cautiously hopeful that LLMs could be used to promote a marginally more balanced footing among decision makers in policy negotiations---particularly benefiting developing countries who face capacity constraints that put them at a disadvantage in negotiations---by assisting with certain tedious, resource-intensive work like analyzing and drafting legal and policy instruments, building capacity, and aiding public consultation processes.
However, LLMs' poorly-understood risks are particularly concerning for environmental and marine policy uses, due to the urgency of crises like climate change, high uncertainty, and trans-boundary impact.

To explore the realistic potentials, limitations, and equity risks for LLMs in marine policymaking, we present a case study of an AI chatbot for the recently adopted Biodiversity Beyond National Jurisdiction Agreement (BBNJ), and critique its answers to key policy questions. The BBNJ Agreement's decades of controversial international negotiations make it a salient case study in equity, and our bot serves as a model of increasingly-popular data-connected AI applications for analyzing such texts.
Our case study demonstrates the dangers of LLMs in marine policymaking via their potential bias towards generating text that favors the perspectives of mainly Western economic centers of power, while neglecting developing countries' viewpoints. 
We describe several ways these biases can enter the system, including: (1) biases in the underlying foundational language models (like GPT); (2) problems arising from the AI chatbot's connection to UN negotiation documents, like disproportionately over-representing affluent viewpoints, or failing to correctly interpret the subtle nuances of polite diplomatic language; and (3) biases arising from the design of the chatbot program (aka. ``prompt engineering''). While we urge caution in the use of generative AI in ocean policy processes and call for more research on its equity and fairness implications, our work also underscores the need for developing countries' policymakers to develop the technical capacity to engage with AI on their own terms.

\end{abstract}

% Use if graphical abstract is present
% \begin{graphicalabstract}
% \includegraphics{figs/grabs.pdf}
% \end{graphicalabstract}

% Research highlights
% \begin{highlights}
% \item Research highlights item 1
% \item Research highlights item 2
% \item Research highlights item 3
% \end{highlights}

% Keywords
% Each keyword is seperated by \sep
\begin{keywords}
AI \sep Large Language Models \sep Marine Policy \sep Environmental Justice
\end{keywords}

\maketitle

\section{Introduction}

Recent breakthroughs in generative artificial intelligence, especially Large Language Models (LLMs) like ChatGPT,~\cite{ouyangTrainingLanguageModels2022,openaiGPT4TechnicalReport2023} allow computer systems an unprecedented (albeit limited) ability to ``understand'' documents in context and engage in natural-feeling conversations about complex topics. These recent LLM breakthroughs have caught researchers by surprise, seemingly exhibiting AI abilities that most thought were still decades away. LLMs are also the subject of massive hype and confusion. For example, pundits speculate that AI will dramatically reshape many industries in a ``fourth industrial revolution''~\cite{french4thIndustrialRevolution2021,mandaRespondingChallengesOpportunities2019, skilton4thIndustrialRevolution2018}; AI has also become one of the most-discussed topics in corporate earnings calls~\cite{ChiefExecutivesCannot2023, OurEarlyadoptersIndex}; and tech startups are awash with venture capital investments as they race to develop experimental LLM-based software for nearly every industry~\cite{LargeCreativeAI2023}.

LLMs are already having an impact on marine policy-making processes, despite their risks being poorly understood. A number of this paper's authors have already observed State representatives and delegates using ChatGPT at the UN for purposes including the drafting of interventions, statements, submissions, and biographies; asking it questions to conduct background research; and even generating whole presentations. Some countries have already developed policies for ChatGPT use for their governmental officials~\cite{UseArtificialIntelligence}. Furthermore, academic researchers are increasingly using ChatGPT to assist with writing articles~\cite{vandisChatGPTFivePriorities2023}, and journalists are utilizing ChatGPT to summarize and interpret official documents~\cite{manjooChatGPTAlreadyChanging2023}.

The best-known LLM application, ChatGPT, uses a general-purpose language model (GPT) to engage in conversations about a wide array of topics with moderate accuracy. However, many researchers and companies are also racing to develop differentiated specific-purpose LLM tools that allow AI to perform better at specific tasks, by connecting LLMs with relevant data or programming abilities to interact with the real world~\cite{chaseLangChain2022} (such as answering questions about a company's internal knowledge base, for example). 
We expect to see these data-connected LLM applications become much more popular in the near future as the tech industry continues to push in this direction: for example, OpenAI (the company behind ChatGPT) has announced plans to open a ``GPT Store'' for customized agents~\cite{openaiIntroducingGPTs2023, metzOpenAILetsMomandPop2023}, Google has launched ``Notebook LM'' for analyzing document collections with AI~\cite{raizamartinIntroducingNotebookLM2023}, and Microsoft is marketing its ``365 Copilot'' that connects GPT to a company's internal documents~\cite{spataroAnnouncingMicrosoft3652023}.

Data-connected AI tools are already finding their way into policy arenas: for example, one can already find commercial and experimental LLM tools for tasks like generating economic impact analysis reports~\cite{woodruffAutomatePolicyAnalysis2023}, summarizing and answering questions about legislation~\cite{dutiaHackingAIClimate2022, angeloEmpoweringPolicyMonitoring2023}, and finding information in large document databases for UN negotiation processes~\cite{UsingAugmentedIntelligence2023}. Efforts in other sectors are slightly more mature; for example, using LLMs to make sense of vast document databases in medicine~\cite{leeBenefitsLimitsRisks2023,sifatChatGPTFutureHealth2023}, law~\cite{basuParalegalsRaceStay2023}, and academic research~\cite{dwivediOpinionPaperWhat2023, vandisChatGPTFivePriorities2023}.

Similar specialized LLM tools are likely to find their way into ocean policy processes in the near future. AI tools are poised to help policymakers with a variety of tedious tasks, like understanding complex legal documents, drafting often-repetitive and formally-worded statements and policy instruments, or quickly finding answers to specific research questions. LLM tools could be especially useful for policymakers in developing countries, who tend to be overstretched and under-resourced compared to their peers in the developed world.

However despite their massive hype and rapid uptake, LLM tools' risks, potential applications, and inner workings are still poorly understood. Researchers are racing to understand how these tools work and why they are so good at what they do, whilst attempting to characterize their emergent behaviors. For example, it is still hotly contested how much LLMs are able to ``reason'' about new problems, versus how much they only parrot sentences and patterns from their vast training data~\cite{liangHolisticEvaluationLanguage2023,metzMicrosoftSaysNew2023,bubeckSparksArtificialGeneral2023,carliniQuantifyingMemorizationNeural2022,ullmanLargeLanguageModels2023}. 

Concerningly, a substantial and growing body of research is documenting inherent biases in popular AI language models. These biases tend to emanate from their training data and design processes, and embody harmful racial and gender stereotypes~\cite{nobleAlgorithmsOppressionHow2018, gonenLipstickPigDebiasing, liangHolisticEvaluationLanguage2023}. Already, activists and researchers have documented real-world consequences of AI biases, such as discrimination in hiring systems for tracking and filtering job applications~\cite{chenEthicsDiscriminationArtificial2023,buylTacklingAlgorithmicDisability2022,sanchez-monederoWhatDoesIt2020}; discrimination in algorithmically-targeted advertisements for work and housing~\cite{kingsleyAuditingDigitalPlatforms2020,lambrechtAlgorithmicBiasEmpirical2019,aliDiscriminationOptimizationHow2019}; and the use of AI for criminal sentencing, supposedly predicting a defendants' risk of re-offending~\cite{parkInjusticeExMachina2019,angwinMachineBiasThere2016,aaronsankinCrimePredictionSoftware2021}.

This article explores the potential equity implications for LLM-based generative AI when it comes to marine policymaking processes. While we worry about the same racial and gender biases that are increasingly documented by critical technology scholars, the distinct and interrelated characteristics of the marine environment present their own set of equity concerns: 
\begin{enumerate}
    \item Marine governance is trans-boundary in nature~\cite{merrieOceanSurprisesTrends2014}.
    \item The marine environment is a highly productive economic space of global consequence with complex dynamics of social inequality at play~\cite{campbellGlobalOceansGovernance2016,clarkExaminingEffectEconomic2019}.
    \item Capacity and power imbalances among oceanic actors remain profound~\cite{tolochkoUsualSuspectsDistribution2021, cisneros-montemayorEnablingConditionsEquitable2021}.
    \item In spite of this, the marine environment is often incorrectly considered to be a non-human space involving no consequential distinctions in terms of culture, race, class, or gender~\cite{gollanItNotJust2020}.
\end{enumerate}    
We argue that, when coupled with the inherent biases and technical limitations of AI language models, the distinctive characteristics of the marine environment create unique risks and considerations for using LLMs in marine policy applications.

\begin{figure*}[h]
    \centering
        \includegraphics[scale=.5,trim={0 5cm 0 0},clip]{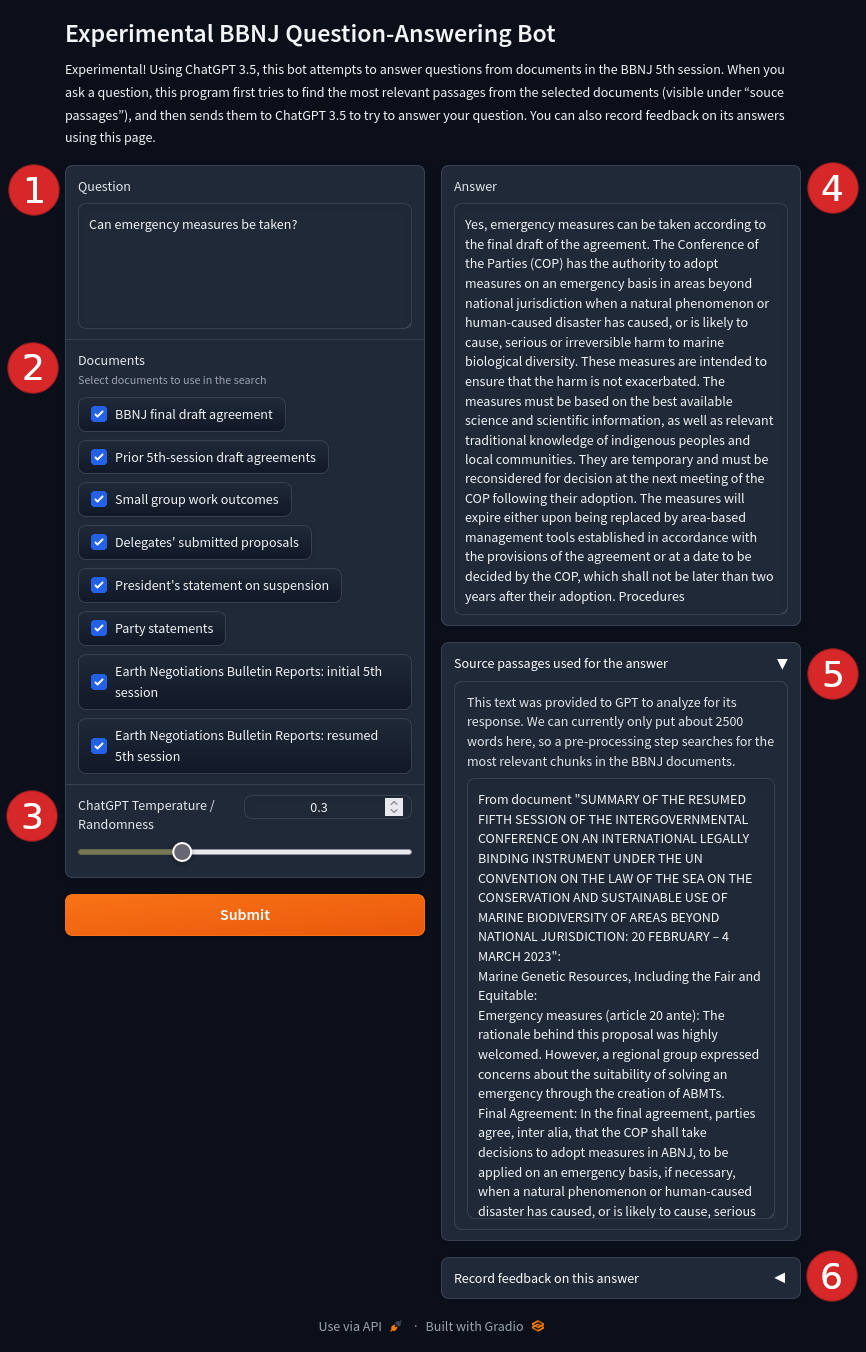}
    \caption[A list enumerating widgets in the screenshot]{User interface screenshot for the BBNJ Question-Answering Bot 
    \begin{enumerate*}
        \item Users type a question related to the BBNJ agreement.
        \item The bot forms its answer after searching for relevant information in a database of BBNJ-related documents. Here, the user can optionally include or exclude some of the sources used for the bot's answer.
        \item A ``temperature'' parameter sent to ChatGPT, that influences the ``randomness'' of the answers it generates.
        \item The bot generates an answer to the user's question and displays it here, after clicking ``submit.''
        \item Here, the user can also browse though the source texts which the bot used to generate its answer. (These are found by searching through the BBNJ documents chosen in step 2, to find the passages most relevant to the user's question.)
    \end{enumerate*}}
    \label{FIG:chatbot-screenshot-1}
\end{figure*}

To evaluate these characteristics and dynamics, this article develops an exploratory case study to examine risks and considerations. We created a \emph{``BBNJ Question-Answering Bot''} (Fig. \ref{FIG:chatbot-screenshot-1}) for the recently adopted \emph{Agreement under the United Nations Convention on the
Law of the Sea on the conservation and sustainable use of
marine biological diversity of areas beyond national
jurisdiction (the BBNJ Agreement)}~\cite{IntergovernmentalConferenceMarine}.
\footnote{The BBNJ Agreement, also known as the High Seas Treaty, is a legally-binding agreement that was adopted by consensus at the United Nations on 19 June 2023 following almost two decades of multilateral negotiations~\cite{ProtectingOceanTime2023}. The BBNJ Agreement opened for signature on 20 September 2023 and will enter into force 120 days after the deposit of the 60\textsuperscript{th} instrument of ratification. The BBNJ Agreement focuses on four thematic issues, namely (1) marine genetic resources including benefit-sharing obligations; (2) area-based management tools including marine protected areas; (3) environmental impact assessments; and (4) capacity-building and the transfer of marine technology.} The BBNJ Agreement makes a salient equity case study for AI tools because (1) it underwent almost two decades of controversial negotiations between blocks of ``developed'' and ``developing'' countries and it has been routinely criticized as a neocolonial policy instrument that serves the interests of wealthy economic centers of power~\cite{tolochkoUsualSuspectsDistribution2021}; and (2) national governments must now determine how to effectively implement the BBNJ Agreement's obligations within their domestic legal systems---a task that is likely to be exceedingly burdensome for many resource-constrained governments, but something LLM tools could be useful for, given their ability to interpret large document sets.

We describe our experimental bot in Section \ref{section:casestudy} and Appendix \ref{appendix:chatbot}, and situate it as a representative model for likely future AI policy tools. We also provide some relevant technical background about LLMs (section \ref{section:technical}) to demystify their workings and explain how their biases and limitations arise. 

We are especially concerned about how AI applications can be biased towards the perspectives of developed countries, and how their use at the UN could further sideline developing countries in negotiation processes. We demonstrate and characterize these biases by analyzing responses from our BBNJ Question-Answering Bot (Section \ref{section:biases}). As one such example, when asked various questions regarding the access and benefit-sharing mechanism of the BBNJ Agreement for marine genetic resources, the chatbot consistently lauds the Agreement's equity accomplishments in a triumphant positive light, while largely ignoring the substantial challenges faced by under-resourced developing countries when it comes to ensuring the mechanism works in practice. 

We outline several ways biases can enter the chatbot system, including (1) biases in training data for the underlying foundational language models (like GPT); (2) problems arising from the AI chatbot's connection to UN negotiation documents, including disproportionate over-representation of affluent viewpoints, as well as the models’ difficulty interpreting the subtle nuances of polite diplomatic language; and (3) biases arising from the design of the chatbot program (aka. "prompt engineering"). We also outline social and institutional factors that could allow these AI errors and biases to perpetuate into policymaking processes, like over-trust and overreliance on AI, potentially de-skilling the workforce and displacing real capacity building.

While urgent attention should be directed towards problems and biases of LLMs, we also remain cautiously hopeful that they could play a role in promoting a more balanced footing among delegations in ocean policy negotiations (Section \ref{section:opportunities}). Especially considering how many developing countries currently lack access to the necessary legal and technical expertise and the financial, human and technological resources to participate in these processes, LLM-based AI tools could provide useful capacity-building assistance when it comes to understanding legal and policy instruments and aid public consultation processes, providing easier access to information. 

We certainly do not expect LLMs to radically upend the power dynamics of the UN, but LLMs are still likely to change how policymakers do their work, similarly to how the adoption of word processing software or the Internet changed policymaking processes. There is still potential for LLMs to empower developing countries in these international forums, if specifically pursued with enough intention and resources.
Furthermore, our work underscores the need for developing countries to build their own technical capacity, to engage with AI on their own terms instead of relying on centralized power from the technology industry and developed countries~\cite{hassanGoverningAlgorithmsSouth2022}.

\section{Case Study: BBNJ Question-Answering Bot}\label{section:casestudy}

We built a \emph{BBNJ Question-Answering Bot} as a case study. This paper explores the potentials, limitations, risks, and equity implications for AI language models in marine policy by examining the bot's construction and question-answering behavior. We will make the souce code for the BBNJ Question-Answering Bot publicly available, along with its database of BBNJ negotiation documents.

The bot is accessed via a web page (Fig. \ref{FIG:chatbot-screenshot-1}). Users type a question about the BBNJ Agreement, and the chatbot responds with an answer. Under the hood, the bot uses a common design pattern for data-connected AI chatbot applications at the time of writing in 2023: ChatGPT provides conversational instruction-following capabilities and a contextual ``understanding of the world,'' but it lacks specific expertise on the BBNJ Agreement so we have connected it to a database of BBNJ-related documents and instruments (Fig. \ref{FIG:application-flow}). When the user types a question, our application first searches for the most-relevant passages from the BBNJ documents, and then sends them to ChatGPT to find an answer. (More technical details are available in Appendix \ref{appendix:chatbot}).

In this paper we consider examples of conversations from our BBNJ Question-Answering bot to analyze its strengths and weaknesses. We posit that findings from our chatbot will be generally applicable to other similar uses of AI language models in policy because it is built using the same predominant design pattern as other LLM applications in policy and elsewhere~\cite{castroRevolutionizeYourEnterprise2023, dutiaHackingAIClimate2022}---including those in OpenAI's forthcoming ``GPT store''~\cite{openaiIntroducingGPTs2023,metzOpenAILetsMomandPop2023}---by connecting the AI model to external data by searching for relevant passages and including them in the model's context window. (Furthermore, we do not have much reason to believe that future technology improvements---like improved foundational models or new application design patterns---would lead to major differences in biases or equity implications.) Methods for evaluating LLM applications are still emerging. Some related work has developed quantitative benchmarks to measure the frequencies of occurrences of biases like racism and sexism~\cite{gonenLipstickPigDebiasing, liangHolisticEvaluationLanguage2023}, but this type of frequency measurement is outside the scope of this work.

Other work has used qualitative methods to characterize harms from LLMs that are more subtle and nuanced, and harder to automatically detect in a quantitative benchmark~\cite{gadirajuWouldnSayOffensive2023}.
Similarly, we use qualitative analysis to offer a preliminary exploration that outlines some of the potential problems with LLM-based tools in environmental policymaking.

\section{Technical background}\label{section:technical}

This section briefly explains some relevant technical background, providing context to our forecast for future LLM policy tools, and how biases and equity issues arise. (As the development of research AI language models is evolving rapidly, one must be cautious about speculating too far on how the landscape will change in the future.)

\subsection{Large Language Models: Training and Text Generation}

At their core, AI language models operate by learning patterns and associations between words from "training" on a massive set of example texts. 
Generative decoder models like GPT produce text via a surprisingly simple process: producing one single word at a time, by guessing the most likely word based on the prior words in a sequence of text.\footnote{The current most powerful and popular LLMs are decoder models (including GPT, Google's PaLM, and Meta's LLaMA) that only consider preceding words when generating text, one word at a time. However, other LLM model architectures exist which generate text in different ways, like Google's "Bidirectional Encoder Representations from Transformers" (BERT) models. It is possible that these prevailing one-word-at-a-time model structures will change in the future.} (This is analogous to the familiar auto-suggest feature on smartphone keyboards, which suggests likely next-words as you type.) Importantly, these AI models have no ability to "plan" their response; they are simply just trying to guess the next word based on the previous words in the sentence, based on patterns observed in their training data.

And yet, despite their simple mechanism of only guessing one word at a time, they are able to learn impressively complex patterns because they represent the associations with enormous models having billions (or even trillions) of crudely-simulated neurons, connected to each other in many layers. The recent LLM boom was caused by a key innovation that allows more-efficient representation of large texts in the models~\cite{vaswaniAttentionAllYou2017a}, allowing them to make associations not only between nearby words in a sentence, but also associating patterns between words and clauses in other paragraphs, and further up and down different parts of the text. These efficiency improvements enabled well-resourced technology companies to train models on massive bodies of text comprising of trillions of words (indeed almost all of the public text on the Internet, among other sources~\cite{brownLanguageModelsAre2020}), teaching the models patterns and associations between many different concepts and incorporating an enormous amount of information; these associations give the model a contextual "understanding of the world." 

LLMs with conversational capabilities, like ChatGPT, are able to follow a user's instructions because the models' training data included many examples of conversations and instruction-following, and gave them special emphasis. For these reasons, applications like ChatGPT are able to have a natural-feeling conversation with a user, even though they are essentially still just trying to guess the next word by imitating patterns in their training data.

\subsection{Connecting to document databases}
Whereas general-purpose LLM tools like ChatGPT can engage on a variety of topics with moderate accuracy, companies and researchers are increasingly building specific-purpose LLM tools that are becoming more accurate and capable of being applied to a specific topic, by connecting them to a data source. For example, we connected our BBNJ Question-Answering Bot to a database of BBNJ-related documents (listed in fig \ref{FIG:chatbot-screenshot-1}); example chats \fakeref{1.1} and \fakeref{1.2} demonstrate how it is able to answer BBNJ-related questions more precisely than regular ChatGPT. There are already a variety of free and commercial products on the market following this design pattern in response to industry demand in different sectors, from analyzing medical records to chatting about a company's internal knowledge base~\cite{castroRevolutionizeYourEnterprise2023, chaseLangChain2022, turnerEpicMicrosoftBring2023, metzOpenAILetsMomandPop2023,openaiIntroducingGPTs2023}.\footnote{
There are currently a few techniques to connect LLMs with document databases. At the time of writing, the most popular method is to first search the database for relevant passages, and simply include them in the prompt (``context window'') that includes the instructions that is processed by the language model. Our BBNJ Question-Answering Bot uses this method (described in Appendix \ref{appendix:chatbot}). This approach has weaknesses in the search step (if the program fails to find the relevant passages in the documents), and context-size limitations (technical restrictions on how many words of text can be processed by the model at a time; currently a few thousand depending on the model).

Other methods for connecting LLMs and document databases are likely to become more popular in the near future as technology improves: in ``fine tuning'' approaches, the general-purpose language model is re-trained on a specific set of documents; and ``plugin'' approaches the LLM is able to trigger programs to look up information on-the-fly. We speculate that AI policy tools using these other methods will produce similar equity issues, but might enable different technical strategies to mitigate them.}

\subsection{User Interaction Designs}
While the most famous LLM application (ChatGPT) is a chat bot, as is our BBNJ Question-Answering Bot, it is worth noting that there are other ways to interact with LLMs that don't involve chat. Many of these alternative interaction styles will likely show up in future AI tools for marine policy. For example, new AI writing-aid software designs use LLMs to offer suggestions and fact-checking as you write~\cite{dutiaHackingAIClimate2022, warrenMicrosoftLookingOpenAI2023, alikaniotisUnreasonableEffectivenessTransformer2019}. Some software designs are centered around particular documents (like legislation or web pages) and offer features to help understand or research different passages in the document~\cite{dutiaHackingAIClimate2022, xiaoInformUninformedImproving2023}. Additionally, software designers are experimenting with more complex workflows to automate complicated analysis of large document databases~\cite{UsingAugmentedIntelligence2023}.

\subsection{Poor Explainability}
LLMs and other AI tools using neural models are plagued by poor ``explainability.'' When an AI agent generates an answer, it is very difficult to trace back its steps and understand why it answered the way that it did because of the models' massive complexities. Improving AI explainability and accountability is currently a hot research topic~\cite{adadiPeekingBlackBoxSurvey2018}, but our poor understanding of their inner workings makes it difficult to trust LLMs for important applications like ocean policy. 

Further exacerbating this problem, GPT and most of the other popular LLMs are ``black boxes,'' such that their model structure, parameters, and exact training data are kept secret~\cite{openaiGPT4TechnicalReport2023}. This makes them even more difficult to scrutinize, as researchers can only study them by examining their behavior.

\subsection{Limitations of Analysis and Reasoning}

LLMs are very good at certain tasks stemming from their abilities to detect patterns; producing fluent-sounding documents and conversations; summarizing complex texts, picking out relevant details, and sometimes relating information from different sources (examples in Appendix \ref{appendix:chats-goodanswers}). However, the extent of their other ``reasoning'' abilities is highly controversial among computer science researchers: some evidence arguably suggests that the models have learned complex reasoning skills due to their massive parameter spaces, some going as far as to claim LLMs exhibit complexities like abstract thought and theory of mind~\cite{bubeckSparksArtificialGeneral2023}. Other work has cast doubt on these findings, arguing instead that the models are just parroting back text from their vast training data instead of deducing the answers themselves~\cite{ullmanLargeLanguageModels2023, carliniQuantifyingMemorizationNeural2022,metzMicrosoftSaysNew2023}.

And so, as even computer science researchers are struggling to detangle the reasoning capabilities of LLMs, this is even more confusing to the general public. When people use LLMs for policy analysis, it can be hard to understand the models' capabilities and limitations for conducting analysis, and gauge which questions are appropriate to ask an AI bot and how to frame those questions. It is hard to know how the AI bot arrived at its answer, and what text may have influenced it from the model's training or the document database (See example chat \fakeref{2.5}).\footnote{This situation may change in the near future as researchers experiment with various ways to have LLMs ``think through'' more complex problems through techniques like ``chain of thought'' prompting or various ways of breaking problems down to networks of smaller sub-problems~\cite{chaseLangChain2022, yaoTreeThoughtsDeliberate2023, weiChainThoughtPrompting2022}.}

\subsection{Errors and Hallucinations}
Another problem plauging LLMs is that they can sometimes produce errors in their answers, and these errors can oftentimes be hard to predict and difficult to detect. Some are errors by omission---i.e. the model simply misses some bit of information (example chat \fakeref{2.4}); these can stem from both weaknesses in the language models or weak application design, i.e. whereby our chatbot's passage-search step might miss some important passages (Section \ref{section:casestudy} and Fig. \ref{FIG:application-flow}).
Another type of common error is popularly known as ``hallucinations.'' These stem from the LLM's mechanism of generating text by guessing one word at a time: they can sometimes generate nonsensical answers that sound plausible but are not rooted in reality (example chats \fakeref{2.1} and \fakeref{2.2}).
\footnote{There are a variety of ongoing research efforts to reduce LLM errors and hallucinations; such as LLMs that error-check their output using a second LLM; or model structures that tie the output closer to the ground-truth input data~\cite{leeBenefitsLimitsRisks2023, manakulSelfCheckGPTZeroResourceBlackBox2023,waltersFabricationErrorsBibliographic2023}} One of the most famous examples of hallucinations is ChatGPT's tendency to make up fake citations when asked to cite its sources; often making up plausible-sounding article titles and author lists, as well as a fake DOI code for an article that doesn't exist~\cite{waltersFabricationErrorsBibliographic2023}. Although the answers are erroneous, these errors and hallucinations can be difficult to detect when the AI bots present wrong information in a confident tone, making them sound very plausible.

\section{Biases and Equity Concerns}\label{section:biases}
This section elaborates our concerns that LLMs could further disadvantage developing states and other marginalized actors in marine policymaking. We explore potential biases in AI models and their applications; and also the social and institutional factors which could perpetuate them, like misplaced trust or over-reliance on AI.

\subsection{Sources of Bias}
By considering example chats from our BBNJ Question-Answering Bot, here we demonstrate its biases and tendencies, wherein we were specifically trying to assess how it might favor the perspectives of "developed,"  or what we would call central economic powers. These mainly appeared through the chatbot as Western advanced-capitalist nation-states. We present example chats and a detailed critique in Appendix \ref{appendix:chats} that highlight equity implications. It is reasonable to expect that these biases would also be present in other AI tools for international policy in general, due to the fact that the bot designed for this research follows one of the most common design patterns for LLM applications today.

\subsubsection{Biases in underlying language models}
Modern chatbot applications are built on top of AI language models to give them a contextual ``understanding of the world.''
Consequently, biases lie within these models themselves. 
The creators of the most popular LLMs (GPT, Bard, and Llama) do not reveal what data the models are trained on, (only that it has vast amounts of text from the Internet, as well as published articles and books). But, as scholars have tried to deduce, it is likely that model training data skews to over-represent the viewpoints of developed countries, as they have produced the most online content. Racial, gender, and nationality biases are well-documented in these models~\cite{venkitNationalityBiasText2023, bansalSurveyBiasFairness2022, nobleAlgorithmsOppressionHow2018}. The secrecy surrounding LLMs' training data and inner workings makes them difficult to research though, and their biases may be worse than researchers realize at present.

In respect to ocean-related issues, producing and accessing relevant information and data tends to be expensive and limited to academic and scientific institutions in developed countries, thus the training is likely to over-represent the perspectives of developed States. For example, in Example Chat \fakeref{3.1}, our BBNJ Question-Answering Bot erroneously answered that the BBNJ Agreement's preamble refers to ``equity'' as a foundational principal when it actually does not.\footnote{Instead, the principle of equity is listed as a guiding principle in Article 7 of the Agreement and should be used by State Parties in achieving the objectives of the instruments.} We suspect that the bot erred here by parroting language in GPT's training data, like the masses of corporate websites that exult equity and its ostensible importance for the company.
Furthermore, since the models learn the strongest representations of speech patterns from the authors used in their training data, we speculate that this bias in language patterns could negatively impact the models' abilities to interact with policy documents written in different voices by developing countries, particularly in small island developing States, though this requires further research.

\subsubsection{Biases in an applications' document database}

Our BBNJ chatbot interacts with a database of legal policy documents (similar to many other AI tools), and this inevitably introduces biases into the system through several different mechanisms.

Oftentimes, developed countries' perspectives are over-represented in UN documents because of entrenched institutional power, an issue that continues to be documented by critical sociologists and others who have conducted ethnographies inside such institutions~\cite{falzonIdealDelegationHow2021,goldmanImperialNatureWorld2005}.  In fact, this general phenomena of international institutions wielding tremendous power and influence over developing countries has been documented at numerous points and persists at the level of global NGOs and their "field offices" that seek to, in sociologist Michael Goldman's words "discipline" in the sense of creating ever new fields of knowledge (e.g., interdisciplinary social sciences), while disciplining subjects into being better market actors~\cite{goldmanBirthDisciplineProducing2001}. Additionally, developed countries have much greater capacity to simply produce statements, peer-reviewed papers, and media articles supporting their positions---in fact, critical political ecologists and development scholars have sometimes wondered if this is all they actually do~\cite{goldmanConstructingEnvironmentalState2001,goldmanHowWaterAll2007,milneCorporateNatureInsider2022}.

 Developing countries generally have less capacity to produce these documents, and their scholars will take much longer to write about the BBNJ's shortcomings from their perspective. It is also possible that they may be so structurally constrained in their activities, that they do not simply have the time to spend producing documentation to, in effect, compete with the developed countries~\cite{hullDocumentsBureaucracy2012}. In fact, many developing countries' positions and issues with the BBNJ Agreement are not well-documented, or simply not translated into English, resulting in a very limited batch of information in the public sphere.

Furthermore, many of the developing countries' issues and complaints go undocumented because of certain organizational cultural codes of distinction, politeness, and decorum at the UN, or what sociologist Fiona McConnell has described as "repertoires of diplomatic behavior"~\cite{mcconnellPerformingDiplomaticDecorum2018,jonesPerformanceEmotionsDiplomacy2019}. 
For example, complaints must be raised discreetly so as not to offend other parties; (e.g. parties will say they ``had very rich discussions'' or ``robust exchanges'' instead of writing: ``we disagree''). For instance, the official documents reviewed as part of this research rather triumphantly celebrate the BBNJ negotiations' successes, but tiptoe around the failures. As evidenced by our bot, GPT struggles to interpret the subtleties and veiled meanings hidden behind the polite tone.

Example Chat \fakeref{3.2} from the BBNJ Question-Answering Bot illustrates both of these problems: the bot triumphantly declares that \emph{``access and benefit sharing have been prioritized in the BBNJ treaty, particularly for developing countries,''} whereas this was actually one of the main controversies of the negotiations, and diplomats from many developing countries would disagree that access and benefit-sharing are actually prioritized in the final Agreement. In this case, the bot most likely arrived at this answer by parroting the negotiation documents' ceremonial nicities that celebrate the Agreement's successes: for example, section 2 of the Agreement is titled \emph{``Marine Genetic Resources, Including The Fair And Equitable Sharing Of Benefits''}~\cite{IntergovernmentalConferenceMarine}; whether it actually accomplishes this is a matter of dispute. On the contrary, many of the developing countries' criticisms of this mechanism are not publicly documented and thus not available for the bot to use in its answer.  

Much of the important behind-the-scenes disagreements never find their way into official documents, and this disproportionately affects  the voice of already marginalized country actors. AI tools relying on these documents could therefore end up further sweeping the key structural issues at play "under the rug".

\subsubsection{Biases in prompting and application design}

Biases can also be introduced by the technical design of a chatbot or other LLM implication; and the AI models' behavior is influenced by subtle distinctions in how the application programmatically interacts with it. A model can generate different answers depending on the cultural or regional vernacular that is used to ask it a question, or the tone of voice; and many users do not realize that different wordings of a semantically-equivalent prompt can yeild dramatically different results~\cite{zamfirescu-pereiraWhyJohnnyCan2023}. For example, our BBNJ Question-Answering Bot gave completely opposite answers to a question worded in different tones (Example Chat \fakeref{3.3}). Since LLM applications commonly generate the prompt from a template (i.e. Box \fakeref{1}), a software designer or user could carefully exploit this behavior to "steer" the model in different ways. (I.e. starting a prompt with ``I'm a conservative'' or ``I'm a liberal'' in a prompt will lead to different answers to a policy question.) However, this behavior could also lead to unintended outcomes, or different model behaviors for users from different cultural backgrounds.

Additionally, LLM applications commonly employ a variety of safety mechanisms to stop the model from producing harmful or offensive responses, and these have sometimes led to discriminatory outcomes.
For example, some efforts to filter out offensive training text has inadvertently led to disproportional censorship of issues affecting minority groups~\cite{dodgeDocumentingLargeWebtext2021}.

\subsection{Misplaced Trust and Overreliance on AI}

There are these errors and biases, but there are also troubling social and institutional phenomena that threaten to exacerbate their harm. Overtrust and overreliance on AI is one such major concern: overreliance happens when people do not know how much to trust an AI agent, causing them to incorrectly or inappropriately accept its recommendations.

Conversations about misplaced trust and overreliance are especially salient and worrisome now, especially for diplomacy and policymaking, as many people and organizations are grappling with AI for the first time amid hype, confusion, and AI's mystification~\cite{manorOpinionChatGPTThreat2023}. Researchers are working to identify factors that can create appropriate levels of trust and over-trust in AI system; the anthromorphism and human-feeling interactions of LLM chatbots can lead to over-estimation of their capabilities~\cite{lockeyReviewTrustArtificial2021,culleyNoteCautionRegarding2013}; and LLM chatbots' improved abilities to provide explanations for their answers are shown to create more trust, even when the explanations are bogus~\cite{lockeyReviewTrustArtificial2021, krollFallacyInscrutability2018}. Positive first impressions also promote trust with an AI agent~\cite{lockeyReviewTrustArtificial2021}, and thus today in the early LLM boom, many people likely hold ChatGPT in high regard after initially seeing its conversational abilities but do not understand its limitations.

Misplaced trust could cause policymakers and others to overrely on LLMs. These biases can be subtle and dangerously hard to detect: for example, people's views can be unknowingly affected when they co-write with opinionated language models~\cite{jakeschCoWritingOpinionatedLanguage2023}. Confirmation bias can influence people not to fact-check the output of LLMs when their answers align with their prior beliefs, and various forms of \emph{``automation bias''} influence people to tend to favor recommendations from AI systems over other sources~\cite{passiOverrelianceAILiterature2022}. Overtrust can also lead to inappropriate uses for AI, (like relying on LLM chatbots for strategic decisionmaking), or asking inappropriate types of questions (like complex analysis or value-laden judgements, which the LLM mostly answers by repeating text patterns in its training data).

\subsection{Displacing real capacity building}
Capacity-building is a key concept in international
environmental policy and it relies predominately on
technology cooperation, programmes of assistance,
collaborative arrangements and partnerships~\cite{vierrosCapacityBuildingTechnology2020}. These
approaches often rely on individuals from developed
States who possess relevant legal, policy and technical
expertise providing assistance to their counterparts in
developing States. For example, this assistance could
include research and monitoring programmes and the
transfer of knowledge. As capacity limitations could
severely constrain the ability of developing States from
fully and effectively participating in and enjoying rights
granted under the BBNJ Agreement, it is conceivable
that these States could overrely on AI LLM models for
assistance, particularly in terms of understanding the
content of the BBNJ Agreement and their obligations
under the instrument. De-skilling of policymakers is a concern if too many people come to rely on AI~\cite{lockeyReviewTrustArtificial2021}, and this could lead to a
situation where AI displaces real capacity-building. Developing States could find the quick and easily-
accessible answers of a chatbot more appealing than
seeking out assistance from neighbouring developed
States. Conversely, developed states could also try to cut back on cooperation or assistance programs, incorrectly arguing that AI is a sufficient replacement.

\section{Opportunities for marginal equity improvements}\label{section:opportunities}

We are still hopeful that LLMs could yield some positive results for developing and under-resourced countries, despite the equity concerns that we have outlined.

The most promising opportunities stem from the gaping capacity imbalances among oceanic actors: developing States have far fewer financial resources, diplomatic staff, and research capacity to advance their agendas at the UN compared to wealthy powerful States. For example, officials in many under-resourced governments are commonly rotated frequently between different assignments, often needing to jump into negotiations where they have little prior experience or background knowledge, and needing to quickly learn the intricacies of complex issues and lengthy and detailed legal instruments. Officials are over-burdened and busy, needing to quickly write ministerial responses and prepare presentations on a smattering of issues. Some of these gaps are well-suited for AI tools, like helping to draft and understand legal and policy instruments, and aiding with public consultation.

AI is likely to change some aspects of the policy-making process, but we have no reason to expect them to fix these fundamental capacity and power imbalances at the UN. Critical technology scholars have argued that technology improvements tend to amplify the power of those who are best positioned to wield them, rather than ``level the playing field'' or ``democratize information''~\cite{toyamaTechnologyAmplifierInternational2011}. Developed states will also increasingly use AI towards their diplomatic goals, and have better technical capacity to exploit it. (In making this forecast, we look towards analogous technology advances like word processing software and the Internet which changed policy-making without upending the power balance.)
Importantly, AI tools are no substitute for other needed measures like enabling developing countries to hire more diplomatic staff and reforming institutional structures to be more inclusive of developing countries~\cite{osterblumOceanEquity2020}.

However, AI language models can still be another tool in the pockets of developing countries. This section outlines some of the most promising opportunities for developing countries to utilize LLMs in marine policy processes. In some cases we expect developing countries to use commercial off-the-shelf AI tools to fill these needs. However, we still call for specific research attention to these applications, especially since the technology industry will continue to cater to the global north while largely neglecting developing countries~\cite{hassanGoverningAlgorithmsSouth2022}. 
Importantly though, all of these potential applications depend on having trustworthy, accurate, and fair AI tools and language models; necessitating further improvements in the technology.

\subsection{Capacity building}

Multilateral environmental initiatives like the BBNJ Agreement are unlikely to succeed without the participation of all countries~\cite{harden-daviesHowCanNew2022}; and capacity building and technology transfer are still necessary for under-resourced governments in developing countries to fulfill their obligations for protecting the high seas~\cite{vierrosCapacityBuildingTechnology2020, cisneros-montemayorEnablingConditionsEquitable2021}. 
Small, under-resourced governments commonly face particular challenges; like staff shortages, difficulty keeping momentum and institutional memory across political changes and successive government restructuring, and reliance on outside technical capacity~\cite{benzakenGoodGovernanceSustainable2022}. Overstretched 

officials often need to juggle many subjects at once and get up to speed on complex issues. This creates a significant disadvantage: officials from developing countries commonly need to jump into negotiations on unfamiliar topics with little time to prepare, often against better-resourced, specialized negotiators from developed States.

The strengths of LLMs could lend themselves well to these problems, helping officials quickly find the information they need to get up to speed. For example, our BBNJ Question-Answering Bot tended to perform well with questions resolving specific details within voluminous documents (Example Chats \fakeref{1.1--1.4}). Search engines like Google and Bing also exemplify this, as they use LLMs to find quick answers to users' questions~\cite{elizabethreidSuperchargingSearchGenerative2023,kellyMicrosoftBringingChatGPT2023}. 

A particular strength of LLM tools is their ability to provide tailored material for each user's knowledge level, rephrased and explained for inexperienced or expert users. Their speed is also an advantage, able to converse about quickly-changing topics: for example, it takes time to produce textbooks and educational materials about developing countries' positions on the BBNJ Agreement and these resources are yet underdeveloped in 2023, but LLMs (provided with up-to-date data) could help unfamiliar people start engaging with these subjects much quicker than waiting for someone to produce educational materials. To some extent, LLM tools could also help organizations retain institutional memory even when staff turnover is high by helping people engage with information from records of prior negotiations

\subsection{Understanding legal and policy documents}

LLMs tools can potentially help users understand complex legal documents, and this could be especially salient for under-resourced governments with staff shortages and without the same access to legal, policy, and scientific expertise~\cite{vierrosCapacityBuildingTechnology2020}.

The current challenge of ratifying and implementing the BBNJ Agreement illustrates this problem, as countries scramble to understand their obligations under the Agreement and how to pass legislation to implement them in their own domestic sphere. This will require a great deal of work, and many different stakeholders within each government need to parse out intricacies from the Agreement's complicated legal language. For example, government officials need to figure out how to implement the Agreement's Clearing-house Mechanism when it comes to marine genetic resources and how the vaguely-defined processes will work. Other officials, will need to understand their country's obligations when it comes to environmental impact assessments under the BBNJ Agreement. Furthermore, fishing representatives might worry whether the Agreement has any provisions that restrict vessel movements. Many officials lack access to legal resources to confidently answer these questions.

The BBNJ Question-Answering Bot exemplifies how LLMs can be useful towards some of these problems, like answering questions about specific details of the Agreement text, or producing simplified summaries of topics. However, there is potential to extend LLM applications to address more of these needs via technology improvements, creative prompt engineering, and combining additional data sources. For example, one could imagine LLM applications that compare different policies, like explaining how the BBNJ Agreement relates to its parent treaty, the 1982 UN Convention on the Law of the Sea; or comparing the BBNJ Agreement's obligations with a country's current laws and regulations. As another example, officials already commonly use software tools to show changes between different versions of a draft text, but future AI tools could provide additional helpful information, particularly in respect to shedding light on the meaning behind those changes.
Future tools could also provide helpful contextual information as a user reads a document, like finding relevant research articles or policy instruments for each section, or maybe even fact-checking.

\subsection{AI-assisted writing tools}
LLM-powered writing assistants could become one of the most important ways that AI reshapes the policymaking process, and developing countries could be especially positioned to benefit from them. Producing formal documents can be a hugely time-consuming and tedious requirement for participating at the UN: government officials spend considerable time writing procedural or ceremonial boilerplate text, or summarizing and rehashing prior texts~\cite{hullDocumentsBureaucracy2012}. 

LLMs are already especially good at this type of writing, like producing summaries or repetitive documents. In addition, it can also be time-consuming and tedious for government officials to write in the formal tones of the UN, especially when there is a language barrier. LLMs are already well-suited to this, commonly used to re-write ideas in more formal language: Popular commercial writing products like Microsoft Word and Grammarly are already incorporating these features via LLMs~\cite{warrenMicrosoftLookingOpenAI2023,kellyMicrosoftBringingChatGPT2023,alikaniotisUnreasonableEffectivenessTransformer2019}.

We are likely to see increasingly creative technology designs for AI writing tools in the near future, as \emph{human-AI interaction} researchers explore new ways for AI agents to co-produce a document in a dance with a human user~\cite{leeEvaluatingHumanLanguageModel2023, yangAIActiveWriter2022}. For example, future AI writing tools will likely guess what a user is trying to do and try to give helpful suggestions, and engage the user in a dialogue about how the final product should look~\cite{heerAgencyAutomationDesigning2019}. There are ripe opportunities to tailor these interactions for policy-making: an AI tool could look up relevant policy instruments or research as you type, or interactively fact-check your writing on the fly.

\subsection{Public consultation}
Ocean policy covers the largest geographic scope on earth, but the negotiations' inclusivity is extremely narrow: only the people who are allowed to be at the UN can participate. So, governments embark on a variety of consultative processes to engage their citizens and understand the public's views, since each government has an interest to serve their own citizens.

LLMs are well-positioned to help with this consultative work because of their abilities to aggregate and sift through massive amounts of text. For decades, companies and political parties have already been using a smattering of AI tools to understand what customers and voters are saying about them in social media, using approaches like keyword extraction and sentiment analysis to gauge the public's attitudes~\cite{birjaliComprehensiveSurveySentiment2021,hristovaDesignMLbasedAI2022}. LLMs, though, enable much richer and more complex analysis of these public expressions though, perhaps allowing governments to quickly learn about new issues as they arise on social media or enabling pseudo-qualitative summarization of the public discourse for policymakers. In addition to passively listening on social media, it's possible that governments could use LLMs to design programs that invite citizens into the public discourse in new and creative ways~\cite{androutsopoulouTransformingCommunicationCitizens2019}.

\subsection{Technical capacity building}
While AI tools could potentially help organizations build capacity to engage in ocean issues, it is also important to invest in building capacity for developing countries to better utilize and control AI technogies for their own prerogatives, instead of being dependent on donors or outsider technologists~\cite{hassanGoverningAlgorithmsSouth2022}.

As LLMs proliferate, technical capacity-building will become more important as policymakers will need to learn appropriate levels of trust and distrust in LLMs, to make the most from them while avoiding potential pitfalls~\cite{petersImportanceDistrustAI2023}. A policymaker with more experience and training will tend to make fewer errors with AI, since people tend to rely on AI more when working on something unfamiliar~\cite{passiOverrelianceAILiterature2022,lockeyReviewTrustArtificial2021}. Similarly, improving AI literacy can have a major impact: one exemplary study showed that clinicians having low AI literacy were seven times more likely to follow an AI agent's recommendations in medical decision-making scenarios~\cite{kimWhenAlgorithmsErr2020}. 

To an extent, developing countries will be able to mitigate some of the biases in the AI tools that they use when they have more control over the technology itself. Technologists are experimenting with techniques at various stages of the AI pipeline~\cite{liSurveyFairnessLarge2023}, like strategically augmenting them with additional training data to reduce biases without negatively affecting the models' performances~\cite{hongEvaluationTargetedDataset2023}. Recent work, though, has shown that the state-of-the-art \emph{``debiasing''} methods sometimes only work on a superficial level and are incomplete, and more research is needed~\cite{gonenLipstickPigDebiasing}.

\section{Discussion and Conclusion}\label{section:conclusion}

AI Language Models are already finding their way into marine policymaking and we expect to see their use increase further. We have demonstrated both risks and opportunities for LLMs to influence equity in marine policymaking in the near- and medium-term futures though our case study of the BBNJ Question-Answering Bot, and by pulling together relevant research on AI and policy. In our manuscript we have also attempted to demystify LLMs for a policymaking audience, and forecast their future. We do not expect AI to fundamentally restructure power balances at the UN but it will likely still change the working lives of policymaking practitioners.

\paragraph{Cautions: Biases, Errors, Overreliance}
Our work demonstrates the harmful potential for LLM biases to disadvantage developing countries in marine policymaking negotiations, and we outline several ways that these biases can enter the system. We urge technologists to research models and AI applications that mitigate these biases against developing countries, similar to ongoing research on other LLM biases like sexism and racism. Meanwhile, we caution policymakers to be aware of these biases as they use LLMs, or encounter other people's text that may have been written with LLMs. 

Furthermore, we caution against overreliance on LLMs and inappropriate uses. For example, LLMs should not be used for decision-making or answering value-laden questions. (Especially nefarious, this can be fuzzy in practice since LLMs can still subtly influence their user via a variety of psychological phenomena~\cite{jakeschCoWritingOpinionatedLanguage2023}.)

\paragraph{More AI Research on Developing Countries' Policymaking Needs}
We expect technologists and researchers to pursue a deluge of new AI tools for policymaking in the coming years, and we call for special attention for developing countries' particular contexts and needs. We cannot expect that AI technologies made primarily for the developed world will serve the developing countries' needs just as well; as they face additional obstacles like capacity gaps and architectural disadvantages in policymaking processes~\cite{vierrosCapacityBuildingTechnology2020, cisneros-montemayorEnablingConditionsEquitable2021,campbellArchitectureAgencyEquity2022}. Furthermore, additional work is required to mitigate biases in AI models that tend to favor developed countries' perspectives.

\paragraph{Developing Countries need AI Technical Capacity}
Our work also highlights the need for policymakers in developing countries to develop their own AI technical capacity. Developing countries should have the agency to develop their own LLM tools that suit their needs~\cite{hassanGoverningAlgorithmsSouth2022}, and improving AI literacy among policymakers would allow them to engage with AI on their own terms. Furthermore, as other countries will also increasingly be using AI in policymaking and marine conservation, developing countries need the technical capacity to push back against AI misuses. It would be appropriate for the technology industry to assist with this technical capacity building, as well as the developed countries where it is centered. 

\paragraph{AI Should Not Replace Real Capacity-Building and Equity Work}
Though there are some ways that AI can assist developing countries with capacity issues, it will not be a solution to power imbalances in environmental policymaking. There is a danger that the developed world will be tempted to cut back on their capacity-building responsibilities, and attempt to rely on cheap technology instead. We emphasize that AI should not replace real capacity-building and equity work in marine policymaking, especially considering the pitfalls outlined in this paper. It is still vitally important for developed countries to devote resources to capacity building and push towards more equitable policymaking processes, especially because every affected country's full participation is crucial if multilateral environmental efforts are to succeed~\cite{vierrosCapacityBuildingTechnology2020, cisneros-montemayorEnablingConditionsEquitable2021,campbellArchitectureAgencyEquity2022}.

\section*{Acknowledgements}
The authors would like to gratefully acknowledge Angelique Pouponneau, Hussain Sinan, and Angela Abolhassani for helpful conversations that shaped the direction of this work. We are also grateful to the Nippon Foundation Ocean Nexus Center for their support.

\pagebreak

\appendix

\section{Example Answers from the BBNJ Question-Answering Bot}\label{appendix:chats}
This appendix shows selected example chats from our case study, the BBNJ Question-Answering Bot, to illustrate the potential capabilities, limitations, and biases of current LLMs for marine policy applications. Our commentary is somewhat speculative by necessity, because we do not know the inner workings of the underlying AI model (GPT). However, we can still analyze its behavior to reason about likely sources of errors and biases.

These example chats were generated from the bot with all of the available source documents included, and GPT's temperature set to 0.3 (Fig \ref{FIG:chatbot-screenshot-1}).

\subsection{Example good answers}\label{appendix:chats-goodanswers}

\begin{tcolorbox}[float*, width=\textwidth, title=Example \#1.1: Generic ChatGPT vs. Specific-Purpose BBNJ ChatBot, colback=chatmulti-colback]
    ~\\
    \textbf{Question:} What are the modalities for sharing the monetary benefits from marine genetic
        resources in the UN BBNJ agreement?

    \begin{tcolorbox}[width=(\textwidth-4mm)/2,nobeforeafter,title=Generic ChatGPT Answer, box align=top, colframe=chatgpt-colframe!60!black]
        \footnotesize
        As of my last knowledge update in September 2021, negotiations regarding the United Nations BBNJ (Biodiversity Beyond National Jurisdiction) agreement were still ongoing, and the final text of the agreement might not have been determined. However, up until that point, discussions around the modalities for sharing monetary benefits from marine genetic resources were a central aspect of the negotiations. The aim was to establish a fair and equitable mechanism for benefit-sharing while promoting the conservation and sustainable use of marine biodiversity in areas beyond national jurisdiction.
        \\

        The BBNJ agreement covers a range of issues related to marine biodiversity in areas beyond national jurisdiction, including the sharing of benefits arising from the utilization of marine genetic resources. Some of the key modalities and principles under discussion included:
        
            1. Common Heritage of Mankind: This principle implies that benefits derived from marine genetic resources should be shared equitably among all countries, regardless of whether they are located in proximity to the resources or not. This approach is intended to prevent a few countries or entities from monopolizing the benefits.
        
            2. Access and Benefit-Sharing (ABS) Mechanism: The agreement was expected to establish a framework for accessing and using marine genetic resources in a way that ensures that benefits generated from their utilization are fairly shared with both the country of origin and the international community.
        
            3. Marine Technology Transfer: The BBNJ negotiations also considered the transfer of marine technology as a form of benefit-sharing. This could involve sharing advanced technology and scientific knowledge related to marine genetic resources with developing countries to enhance their capacity to study and utilize these resources sustainably.
        
            4. Monetary Benefits: Discussions were ongoing about how to distribute monetary benefits generated from the commercialization of marine genetic resources. Potential approaches included setting up an international fund that would collect royalties or fees from entities benefiting from these resources. The fund could then be used to support marine conservation, research, and capacity-building efforts, particularly in developing countries.
        
            5. Capacity Building and Technology Transfer: Another aspect of benefit-sharing involved enhancing the scientific and technical capacities of developing countries to study, understand, and manage marine genetic resources. This could be achieved through technology transfer, training programs, and collaborations.
        
            6. Sustainable Use: The agreement was expected to emphasize the importance of sustainable use and conservation of marine genetic resources. Benefit-sharing mechanisms would likely take into account efforts to ensure that resource utilization does not harm the marine environment or the long-term viability of the resources.
        \\

        Please note that the information provided is based on the state of negotiations up until September 2021. For the most up-to-date information on the UN BBNJ agreement and its provisions regarding the sharing of monetary benefits from marine genetic resources, I recommend checking official UN sources or other reputable international organizations involved in the negotiations.
    \end{tcolorbox}\hfill
    \begin{tcolorbox}[width=(\textwidth-4mm)/2,title=BBNJ Question-Answering Bot Answer, box align=top, nobeforeafter]

        In the final draft of the UN BBNJ agreement, the modalities for sharing monetary benefits from marine genetic resources include milestone payments, royalties, and other forms as determined by the Conference of the Parties (COP) based on recommendations from the access and benefit-sharing mechanism. The COP is responsible for determining the rate of payments related to monetary benefits. The payments are made through a special fund established under Article 52 of the agreement. The modalities may also include payments or contributions related to the commercialization of products, including a percentage of the revenue from sales of products, and a tiered fee paid on a periodic basis. If all efforts to reach consensus have been
        exhausted, a decision shall be adopted by a three-fourths majority of Parties present and voting.

        \tcblower
        ~\\
        \textbf{Commentary:} This example shows the usefulness of a specific-purpose chatbot application, compared to using a general-purpose chatbot like ChatGPT. (The BBNJ Question-Answering Bot uses GPT 3.5, the same underlying AI language model as ChatGPT, but has passages from BBNJ-related documents loaded into its context window.)
        \\

        In the answer from generic ChatGPT, it demonstrates that it still has some ``knowledge'' about the BBNJ negotiation process, and its answer lists many related topics that arose in discussions. However, the final text of the Agreement is not included because it was completed after ChatGPT's ``knowledge cutoff'' of September 2021. Its answer also is nonspecific and rambles, sometimes venturing into unrelated topics like technology transfer.
        \\

        The BBNJ Question-Answering Chatbot's answer is grounded in the text of the Agreement and other official documents from the BBNJ negotiation process: it is able to give a correct and fairly strong summary, and cites relevant sections of the Agreement including Article 52 which relates to the financial mechanism to be established under the new instrument.
    \end{tcolorbox}

\end{tcolorbox}

\begin{tcolorbox}[float*=h, width=\textwidth, title=Example \#1.2: Generic ChatGPT vs. Specific-Purpose BBNJ ChatBot, colback=chatmulti-colback]
    ~\\~\\
    \textbf{Question:} Does the BBNJ agreement apply to digital sequence information?
    \\

    \begin{tcolorbox}[width=(\textwidth-4mm)/2,nobeforeafter,title=Generic ChatGPT Answer, box align=top, colframe=chatgpt-colframe!60!black]
    As of my last knowledge update in September 2021, the BBNJ (Biodiversity Beyond National Jurisdiction) agreement primarily focuses on the conservation and sustainable use of marine biodiversity in areas beyond national jurisdiction, such as the high seas. It was being negotiated under the United Nations Convention on the Law of the Sea (UNCLOS) framework.
    \\ \\
    Digital Sequence Information (DSI) refers to the genetic information found in DNA and RNA sequences, often used in biological research and applications like biotechnology and pharmaceutical development. The regulation of DSI falls under various international agreements related to biodiversity, such as the Nagoya Protocol on Access to Genetic Resources and the Fair and Equitable Sharing of Benefits Arising from their Utilization, which is a supplementary agreement to the Convention on Biological Diversity (CBD).
    \\ \\
    However, the situation may have evolved since my last update, and new agreements or developments may have occurred regarding the regulation of DSI under the BBNJ agreement or other relevant international frameworks. I recommend checking with more recent and authoritative sources to get the most up-to-date information on this topic.
    \end{tcolorbox}\hfill
    \begin{tcolorbox}[width=(\textwidth-4mm)/2,nobeforeafter,title=BBNJ Question-Answering Bot Answer, box align=top]
    Yes, the BBNJ final draft agreement applies to digital sequence information (DSI) on marine genetic resources of areas beyond national jurisdiction (ABNJ). The draft agreement recognizes that access to and utilization of DSI on marine genetic resources, together with the fair and equitable sharing of benefits arising from its utilization, contribute to research and innovation and to the general objective of the agreement. The draft agreement also includes provisions for the notification, management, and sharing of DSI and other associated data and information related to marine genetic resources of ABNJ.

    \tcblower
    \textbf{Commentary:} This example also illustrates the usefulness of specific-purpose LLM applications (like the BBNJ Question-Answering Bot) over generic applications (like ChatGPT). The BBNJ Bot is connected to up-to-date information that is missing from ChatGPT's training data; and furthermore produces a more focused answer by having the BBNJ documents loaded into its context window. Here, the BBNJ Bot gives a correct answer summarizing relevant information from the BBNJ Agreement, while ChatGPT gives a nonspecific answer that meanders around some related topics and cautions that its information is out of date.

    \end{tcolorbox}

\end{tcolorbox}

\begin{tcolorbox}[float, title=Example \#1.3: Good answer]
    \textbf{Question:} What is the process for establishing a high seas marine protected area?
    \\

    \textbf{BBNJ Question-Answering Bot:} Based on the information provided, the process for establishing a high seas marine protected area involves identifying areas requiring protection through the establishment of area-based management tools, including marine protected areas, on the basis of the best available science and scientific information, as well as relevant traditional knowledge of indigenous peoples and local communities, taking into account the application of precautionary principle and an ecosystem approach. Proposals regarding the establishment of area-based management tools, including marine protected areas, under this Part shall be submitted by Parties, individually or collectively, to the secretariat. Parties shall collaborate and consult, as appropriate, with relevant stakeholders, including States and global, regional, subregional and sectoral bodies, as well as civil society, the scientific community, the private sector, Indigenous Peoples and local communities, for the development of proposals, as set out in this Part. The Conference of the Parties shall take decisions on the establishment of area-based management tools, including marine protected areas, and related measures on the basis of the final proposal and, in particular, the draft management plan, taking into account the contributions and recommendations received during the consultation process established under this Part.
    \tcblower
    \textbf{Commentary:} This is a strong answer; the application has accurately found relevant information from the BBNJ Agreement and related documents, and summarized it nicely. In general, the bot gives its strongest answers for these types of questions that ask for a specific detail that can be found in the text its' provided with.
\end{tcolorbox}

\begin{tcolorbox}[float, title=Example \#1.4: Good answer]
    \textbf{Q:} What are the required steps for an environmental impact assessment?
    \\

    \textbf{BBNJ Question-Answering Bot:} According to the various documents, the required steps for an environmental impact
    assessment include:

    1. Screening to determine whether an environmental impact assessment is required.

    2. Scoping to define the scope of the environmental impact assessment.

    3. Conducting the environmental impact assessment, which includes:

    \quad a. Describing the planned activity, including its location.

    \quad b. Conducting a baseline assessment of the marine environment likely to be affected.

    \quad c. Describing potential impacts, including potential cumulative impacts and impacts in
    areas within national jurisdiction.

    \quad d. Describing potential prevention, mitigation, and management measures.

    \quad e. Identifying uncertainties and gaps in knowledge.

    \quad f. Describing the public consultation process.

    \quad g. Considering reasonable alternatives to the planned activity.

    \quad h. Describing follow-up actions, including an environmental management plan.

    4. Preparation of an environmental impact assessment report.

    5. Scientific and Technical Body review of the environmental impact assessment report.

    6. Publication of the environmental impact assessment report and its findings through the
    clearing-house mechanism.

    7. Development of additional guidelines and best practices by the Scientific and Technical
    Body and Conference of the Parties.
    \tcblower
    \textbf{Commentary:} In this example the bot provides a strong summary of the main steps for conducting an environmental impact assessment as per the BBNJ Agreement. This serves as another example of LLMs' strength for finding specific details and summarizing information, especially when connected to a database of relevant BBNJ documents. 
\end{tcolorbox}

\FloatBarrier
\pagebreak
\subsection{Example wrong answers and hallucinations}

\begin{tcolorbox}[float, title=Example \#2.1: Hallucination on controversial issue\, possibly from missing context]
    \textbf{Q:} Are marine genetic resources the common heritage of mankind?
    \\

    \textbf{Bot Answer:} Yes, marine genetic resources of areas beyond national jurisdiction are considered the
    common heritage of mankind and no state shall claim or exercise sovereignty or sovereign
    rights over them. This is stated in multiple paragraphs, including in Article 4 of the Draft
    Agreement, which states that no state shall claim or exercise sovereignty or sovereign rights
    over marine genetic resources of areas beyond national jurisdiction. Additionally, Article 5
    of the same document states that activities with respect to marine genetic resources of
    areas beyond national jurisdiction are in the interests of all states and for the benefit of all
    humanity.
    \tcblower
    ~\\
    \textbf{Commentary:} This answer is incorrect. The BBNJ agreement does not explicitly state that marine genetic resources of areas beyond national jurisdiction are the common heritage of mankind; the agreement did not go that far. Additionally, Article 4 does not relate to this question.
    \\
    \\
    The bot may have given this wrong answer because it is missing context, and does not ``know'' that the \emph{``common heritage of mankind''} refers to a specific legal principle. Rather, it answers affirmatively because it finds similar words and phrases like \emph{``benefit of all humanity.''}
\end{tcolorbox}

\begin{tcolorbox}[float, title=Example \#2.2: Confusing wording due to hallucination]
    \textbf{Question:} Does the BBNJ final draft agreement apply to warships?
    \\

    \textbf{Bot Answer:} Yes, the final draft agreement of the BBNJ does not apply to warships, military aircraft, or
    naval auxiliary. This exception is stated in various documents, including...
    \tcblower
    \textbf{Commentary:} This answer is mostly correct, but confusing: the chatbot initially begins by answering affirmatively "yes," but continues on (correctly) to say that the agreement does not apply. The confusing wording here could easily be misunderstood.
\end{tcolorbox}

\begin{tcolorbox}[float, title=Example \#2.3: Confusion between different draft versions of the agreement]
    \textbf{Question:} How does the BBNJ Agreement define ‘cumulative impacts’?
    \\

    \textbf{Bot Answer:} The BBNJ Agreement offers multiple options for the definition of cumulative impacts.
    Option A defines cumulative impacts as the incremental effects of a proposed activity under
    the jurisdiction and control of a Party when added to the impacts of past, present and
    reasonably foreseeable activities, or from the repetition of similar activities over time,
    including climate change, ocean acidification and possible transboundary impacts,
    regardless of whether the Party exercises jurisdiction or control over those other activities.
    Option B defines cumulative impacts as impacts on the same ecosystems resulting from
    different activities, including past, present or reasonably foreseeable activities, or from the
    repetition of similar activities over time, including climate change, ocean acidification and
    related impacts.
    \tcblower
    \textbf{Commentary:} This answer is wrong, as it refers back to definitions discussed earlier in the negotiations instead of the definition in the final agreement. This illustrates a likely common problem in LLM applications where the model has trouble discriminating between different versions of a document, or between similar documents.
\end{tcolorbox}

\begin{tcolorbox}[float, title=Example \#2.4: Missing information]
    \textbf{Question:} How many members will comprise the access and benefit sharing committee?
    \\

    \textbf{Bot Answer:} According to the information provided, the number of members that will comprise the
    access and benefit-sharing committee is not specified in the documents. However, it is
    mentioned that the committee shall be composed of members possessing appropriate
    qualifications who are nominated by Parties and elected by the Conference of the Parties,
    taking into account gender balance and equitable geographic distribution, and providing for
    representation on the mechanism from developing States, including the least developed
    countries and small island developing States. The terms of reference and modalities for the
    operation of the mechanism shall be determined by the Conference of the Parties.
    \tcblower
    ~\\
    \textbf{Commentary:} This example shows how the bot can sometimes simply miss information. There will be fifteen members, and this is stated in Article 15 of the BBNJ Agreement. This was included in the passages provided to the language model for this question, and it is unclear why the model did not recognize it.
\end{tcolorbox}

\begin{tcolorbox}[float, title=Example \#2.5: Inability to analyze]
    \textbf{Q:} How does the agreement balance single species conservation vs biodiversity?
    \\

    \textbf{Bot Answer:} The provided information does not make it clear whether the agreement prioritizes single species conservation over biodiversity conservation. However, the final agreement aims to ensure the conservation and sustainable use of marine biological diversity of areas beyond national jurisdiction through effective implementation of the Convention and further international cooperation and coordination. Additionally, the objectives of the agreement include, among others, conserving and sustainably using areas requiring protection, protecting and restoring biodiversity and ecosystems, and supporting socio-economic objectives. Therefore, it suggests that the agreement aims to strike a balance between conservation of single species and the broader goal of conserving and sustaining biodiversity in areas beyond national jurisdiction.
    \tcblower
    \textbf{Commentary:} This example shows the a limitation of the bot's ability to analyze and reason: the correct answer is that the BBNJ agreement prioritizes biodiversity conservation. ("Biodiversity" is even in the name of the agreement.) The bot is unable to deduce this and instead it rambles in no particular direction.\\ \\This example also underscores how hard it is for users to discern when a bot is doing its own analysis or being influenced by something from its training data or document database. If there were a sentence in the model training data or BBNJ documents that directly answered this question, the bot would likely have just found it and confidently repeated it as its own answer; which it did not have in this case.\\ \\This question asked the bot to make a value judgement, and its failure also underscores how it's hard to know where the bots' values come from in such a case, or how it would be doing such an analysis. Should it count the occurrences of "biodiversity" vs "species?" Should it look for strong language around "biodiversity" and "species?" Or something else?
\end{tcolorbox}

\pagebreak
\FloatBarrier
\subsection{Example biased answers}

\begin{tcolorbox}[float, title=Example \#3.1: Wrong answer (probably) due to bias in underlying language model]
    \textbf{Q:} Is equity a foundational concept for the BBNJ Agreement?

    \textbf{Bot Answer:} Yes, equity is a foundational concept for the BBNJ Agreement. It is mentioned as one of the guiding principles in the preamble of the agreement, and there are discussions and proposals regarding fair and equitable benefit-sharing throughout the negotiation process. The draft text of the agreement includes options for fair and equitable benefit-sharing, and delegates from developing and developed countries have diverging opinions on the matter.
    \tcblower
    \textbf{Commentary:} This answer is weak. The word equity only appears once in the BBNJ Agreement, but it’s not in the preamble, it is in Article 7 relating to ‘General Principles and Approaches.’ In our application, we are not supplying any documents which celebrate equity as a foundational concept (though there is some ambiguity since some of the documents talk about various aspects of fairness). We suspect that in this case, the chatbot is saying equity is a foundational concept for the BBNJ because GPT is parroting language from its training data; for example, it is likely influenced by the tons of online text on corporate websites that speaks positively about equity and its importance for the company.
\end{tcolorbox}

\begin{tcolorbox}[float*=h, width=\textwidth, title=Example \#3.2: Biases Stemming from Official BBNJ Negotiation Documents, colback=chatmulti-colback, left=0mm,right=0mm, right skip=0mm, left skip=0mm]

    \small

    \begin{tcolorbox}[width=(\textwidth-2mm)/2,nobeforeafter,title=Version 1, box align=top, left=3mm, right=3mm]

        \textbf{Question:} Have access and benefit sharing been prioritized in the BBNJ treaty?
        
        \vspace{2mm}
        \textbf{BBNJ Question-Answering Bot:} Yes, access and benefit-sharing have been prioritized in the BBNJ treaty.\textsuperscript{1} The final draft of the agreement establishes an access and benefit-sharing mechanism, which is composed of members elected by the Conference of the Parties. The mechanism aims to provide guidelines for benefit-sharing, ensure transparency, and facilitate a fair and equitable sharing of both monetary and non-monetary benefits. The agreement also includes provisions for monitoring and transparency, with the access and benefit-sharing committee preparing reports based on information received through the clearing-house mechanism and making it available to Parties for consideration. The Conference of the Parties can determine appropriate guidelines for the implementation of access and benefit-sharing, taking into account national capabilities and circumstances.
    
    \end{tcolorbox}\hfill
    \begin{tcolorbox}[width=(\textwidth-2mm)/2,nobeforeafter,title=Version 2: ``Developing Countries' Perspectives'', box align=top, left=2mm, right=2mm]
    
        \textbf{Question:} From the perspective of developing countries, have access and benefit sharing been prioritized in the BBNJ treaty?
        
        \vspace{2mm}
        
        \textbf{BBNJ Question-Answering Bot} Yes, access and benefit sharing have been prioritized in the BBNJ treaty, particularly for developing countries.\textsuperscript{2} The treaty includes provisions for fair and equitable sharing of benefits arising from marine genetic resources of areas beyond national jurisdiction, and capacity building and transfer of marine technology to developing states parties, particularly the least developed countries and landlocked developing countries. The treaty also establishes a financial mechanism aimed at ensuring efficient access to funding through simplified approval procedures and enhanced readiness of support for developing states parties, taking into account their specific needs and special circumstances. The GEF\textsuperscript{3} has also set aside funding to support national ratification and implementation of the treaty, particularly for developing countries.

    \end{tcolorbox}

    \begin{tcolorbox}[width=(\textwidth-2mm)/2,nobeforeafter,title=Version 3: Drawing out specific viewpoints, box align=top]

        \textbf{Question:}What were Pacific Island Developing States' suggestions for improving access and benefit sharing, and what is their reasoning? What types of financial and non-financial elements do they advocate for?

        \vspace{2mm}
        \textbf{BBNJ Question-Answering Bot:} The Pacific Island Developing States (P-SIDS) suggested several improvements for access and benefit sharing in the final draft of the agreement. They called for a more comprehensive and ambitious access and benefit sharing regime that goes beyond conservation and sustainable use. 
        
        \vspace{2mm}
        In terms of financial elements, P-SIDS advocated for access to funding based on the needs of developing States Parties. They emphasized the importance of considering the needs for assistance of Parties with special requirements, including the least developed countries, landlocked developing countries, geographically disadvantaged States, small island developing States, coastal African States, archipelagic States, and developing middle-income countries. P-SIDS also called for simplified application and approval procedures to ensure efficient access to funding and enhanced readiness of support for developing States Parties.
        
        \vspace{2mm}
        In terms of non-financial elements, P-SIDS emphasized the importance of capacity-building and the transfer of marine technology. They urged international organizations to grant preferential treatment to developing States Parties, particularly the least developed countries, landlocked developing countries, and small island developing States. P-SIDS also highlighted the need to consider the specific needs and special circumstances of developing States Parties in the allocation of appropriate funds, technical assistance, and the utilization of specialized services for the conservation and sustainable use of marine biological diversity of areas beyond national jurisdiction.

    \end{tcolorbox}\hfill
    \begin{tcolorbox}[width=(\textwidth-2mm)/2,nobeforeafter,box align=top, skin=empty, left=2mm, right=2mm]
        \vspace{2mm}
        \textbf{Commentary:}
            These examples illustrate biases that stem from the official documents from the BBNJ negotiations. Most negotiators from developing countries would disagree with the bot's assertion that the BBNJ prioritizes access and benefit-sharing (notes 1 and 2), instead contending that the final agreement does not do enough. However, the bot's triumphant answer echoes the tone of UN documents that celebrate the agreements' achievements---it misses all of the context of the negotiation processes, and the grievances of the developing countries, which are largely undocumented because of the UN's polite etiquette.
            \\

            This series of questions also shows how the bot can be partially steered towards different perspectives by careful wording of the questions. In Version 2 which specifically asks for developing countries' perspectives, it gives some more details on developing countries' demands, but still only depicts them in a positive light instead of describing the shortcomings. These answers would be stronger if the bot directly referenced developing countries' own statements: Version 3 shows how the bot can be coaxed more towards the viewpoints of less-represented parties by using increasingly specific language in the questions.
            \\

            Additionally, the bot's congratulatory reference to the Global Economic Fund (GEF; note 3) demonstrates a bias from the model lacking important context: the GEF is highly controversial among developing countries, as its funding comes with many challenging conditions attached. (Furthermore, each country can only get one grant from the GEF regardless of if it supports the BBNJ agreement or something else, making it unhelpful for most developing countries towards BBNJ implementation.) Many of developing countries' problems with the GEF are not documented---nor are many other important obstacles for developing countries in international politics---so AI chatbots are not ``aware'' of them.

    \end{tcolorbox}
\end{tcolorbox}

\begin{tcolorbox}[float*=h, width=\textwidth, title=Example \#3.3: Opposite Responses from Question Wording, colback=chatmulti-colback, left=0mm,right=0mm, right skip=0mm, left skip=0mm]

    \begin{tcolorbox}[width=(\textwidth-2mm)/2,nobeforeafter,title=Version 1: Strong tone and confrontational answer, box align=top, left=3mm, right=3mm]    
        \textbf{Question:} The BBNJ's process for environmental impact assessments is a failure of the UN to protect developing countries. Please explain its shortcomings.

        \vspace{2mm}
        \textbf{BBNJ Question-Answering Bot:} Based on the information provided, it is not accurate to say that the BBNJ's process for environmental impact assessments is a failure of the UN to protect developing countries. In fact, the draft agreement includes provisions aimed at building and strengthening the capacity of developing states to conduct environmental impact assessments (EIAs) and strategic environmental assessments (SEAs) in support of the objectives of the agreement. This is particularly important for developing countries, including small island developing states (SIDS), which often face disproportionate impacts from activities in areas beyond national jurisdiction (ABNJ).

        \vspace{2mm}
        The draft agreement recognizes the need to achieve a coherent environmental impact assessment framework for activities in ABNJ and includes provisions for conducting EIAs and SEAs. It also emphasizes the importance of considering cumulative and transboundary impacts, as well as the use of the best available science and relevant traditional knowledge of indigenous peoples and local communities in the assessment process.
        
        \vspace{2mm}
        Furthermore, the draft agreement acknowledges the need for capacity building and technology transfer to assist developing states in preparing, conducting, and evaluating EIAs and SEAs. This recognizes the challenges faced by developing countries, including LDCs, LLDCs, and SIDS, and aims to ensure their meaningful participation in the implementation of the agreement.
        
        \vspace{2mm}
        While there may be differing views and discussions on specific aspects of the process for environmental impact assessments, it is important to note that the draft agreement reflects efforts to address the needs and concerns of developing countries and promote their effective participation in protecting biodiversity in ABNJ.

    \end{tcolorbox}\hfill
    \begin{tcolorbox}[width=(\textwidth-2mm)/2,nobeforeafter,title=Version 2: Softer tone and concurring answer, box align=top, left=2mm, right=2mm]   
        \textbf{Question:} From the perspective of developing countries, what are the criticisms of the BBNJ's environmental impact assessment?

        \vspace{2mm}
        \textbf{BBNJ Question-Answering Bot:} From the perspective of developing countries, there are several criticisms of the BBNJ's environmental impact assessment (EIA) provisions. One criticism is that the EIA requirements may impose a burden on developing countries, particularly those with limited capacity and resources. Developing countries argue that they may lack the technical expertise and financial resources to conduct comprehensive EIAs for activities in areas beyond national jurisdiction (ABNJ). This could hinder their ability to fully participate in and benefit from the conservation and sustainable use of marine biodiversity in ABNJ.

        \vspace{2mm}
        Another criticism is that the EIA provisions may not adequately address the specific needs and circumstances of developing countries. Developing countries argue that the EIA requirements should take into account their unique vulnerabilities, such as their dependence on marine resources for livelihoods and their limited ability to adapt to environmental changes. They emphasize the need for capacity building and technology transfer to support their efforts in conducting EIAs and implementing effective measures to prevent significant adverse impacts on marine biodiversity in ABNJ.
        
        \vspace{2mm}
        Additionally, some developing countries express concerns about the internationalization of the EIA review process. They argue that decisions regarding significant adverse impacts and mitigation measures should be made by the relevant/involved party rather than the scientific and technical body. Developing countries are cautious about potential interference in their decision-making processes and the potential for halting activities without their consent.
        
        \vspace{2mm}
        Overall, developing countries seek a balanced approach to the EIA provisions of the BBNJ agreement, one that recognizes their specific needs and capacities, ensures adequate support for capacity building and technology transfer, and avoids duplication with existing IFBs.
    \end{tcolorbox}

    \vspace{2mm}
    \textbf{Commentary: }Here we see different wordings of substantially the same question produce polar opposite answers from the BBNJ Bot (covering nearly the same topics). When asked about the BBNJ's failures in a somewhat confrontational tone, the bot responds defiantly and defends the BBNJ's merits; but it willingly describes criticisms when asked in a softer tone.

    \vspace{2mm}
    This example shows both how bias in LLM applications can come from the user's choice of words, but also from the application designer who mediates how the user's words interact with the underlying AI language model. For example, the bot's opposing perspectives can be caused by the program finding different document passages for each question in the preprocessing-search step (Fig \ref{FIG:application-flow}). Additionally, LLM chatbots are known to respond confrontationally to confrontational prompts~\cite{morrisMicrosoftNewBing2023,billyperrigoBingAIThreatening2023} and the chatbot may have taken a defensive position simply in response to the tone of the question; AI application designers can also thereby cause bias in their applications from their handling or augmentation of the prompt (Box 1).
\end{tcolorbox}

\FloatBarrier

\section{BBNJ Question-Answering Bot: Technical Design and Implementation}\label{appendix:chatbot}

\begin{figure*}[H]
	\centering
		\includegraphics[width=0.95\textwidth]{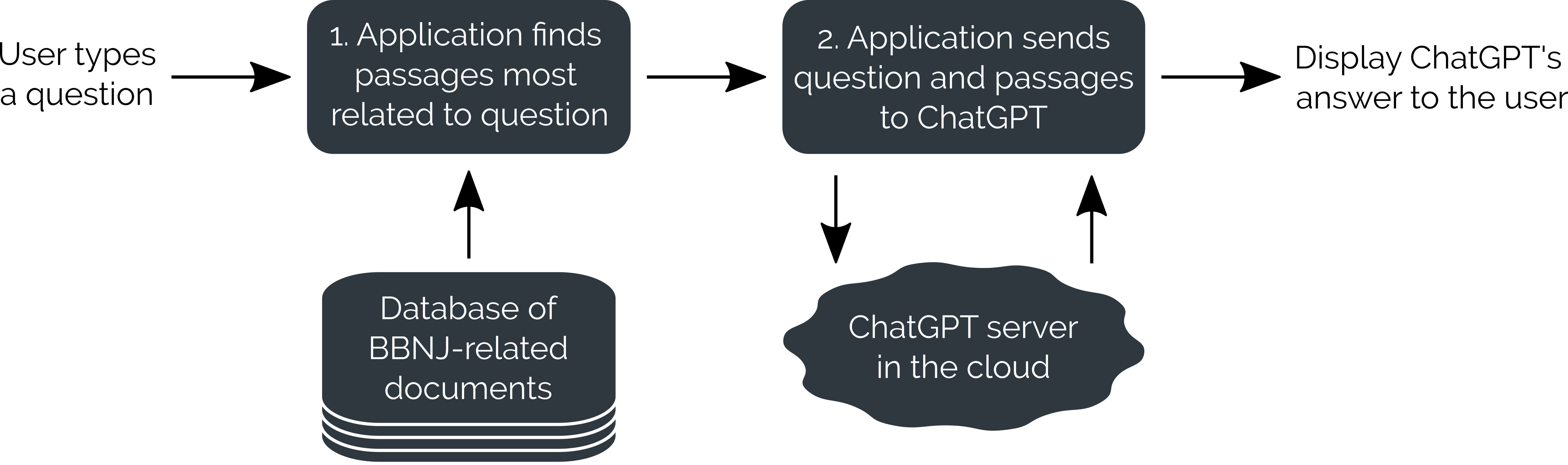}
	\caption{This flow diagram represents the structure of the BBNJ Question-Answering Bot application. When the user answers a question, the program first searches for relevant passages in a database of BBNJ-related documents. It sends the users' question along with the passages to a ChatGPT server in the cloud, and then displays ChatGPT's reply. At the time of writing in 2023, this is a very common design pattern for AI applications that interact with external documents.}
	\label{FIG:application-flow}
\end{figure*}

Our BBNJ Question-Answering Bot follows a common software design pattern in 2023 used to connect LLMs like GPT to a database of specific documents, by incorporating a search step that first finds relevant text passages; and then inserts them into GPT's context window by including them in a ``user'' message, along with the user's question (Fig \ref{FIG:application-flow}). Upon publication of this manuscript, we will make the source code of the chatbot publicly available.

To scope the document database accessed by the chatbot, we decided to include documents from the fifth (final) session of the BBNJ negotiations~\cite{unitednationsgeneralassemblyFifthSubstantiveSession2023}, including:
\begin{itemize}
    \item The BBNJ Agreement finalized text,
    \item Prior draft agreements from the 5th session,
    \item Small-group work outcomes,
    \item Delegates' submitted proposals,
    \item The statement by the President of the Intergovernmental Conference on the suspension of the 5th substantive session,
    \item Party statements, and 
    \item Earth Negotiations Bulletin Reports from both rounds of the 5th substantive session.
\end{itemize}

We used the Science Parse software~\cite{ai2ScienceParse2018} to extract paragraphs from the PDFs, preserving the heading structure when possible. We then separated the text into passages for indexing in an embeddings model, generally following OpenAI's recommendations to index the text as one-paragraph-sized passages~\cite{openaiEmbeddingsOpenAIPlatform2023}. (Through trial and error, we made two modifications to suit the UN documents' formatting, accommodating small document sections by including them as a whole passage when they had under 200 tokens, roughly 150 words. Because the UN documents also frequently used bulleted lists which Science Parse split into paragraphs, we merged together adjacent ``paragraphs'' within the same section, into passages having at least 100 tokens, roughly 75 words.) 

We used an embeddings model to make the passages searchable, a common AI technique that assigns a numeric vector that represents the ``meaning'' of each passage relative to a language model. We generated embeddings with OpenAI's latest \emph{text-embedding-ada-002} model~\cite{openaiEmbeddingsOpenAIPlatform2023}, and stored the resulting vectors in a Weaviate vector index~\cite{dilockerWeaviate2023}. 

Upon receiving a question from the user, our application first searches for the most relevant passages in the vector index to the user's question, (ranked by the lowest angular/cosine distances from each passage to the embedding vector for the user's question). The resulting passages are then flattened/concatenated into a text string to include in the prompt to GPT, as many as can fit into the context window, along with their document titles and passage headings. For the example chats shown in this paper, we allowed up to 3000 tokens (~2250 words) for the included passages, to also allow room for the user's questions, prompt instructions, and generated answer; while staying under GPT 3.5's original context size limit of 4,097 tokens. 

The application then generates an answer to the user's question by sending a request to the GPT API for completion, with the passages, question, and additional instructions embedded into a prompt (Box \fakeref{1}). Upon receiving from the GPT service, the bot application displays the answer to the user. The graphical interface was built using the Gradio library~\cite{abid2019gradio}.

For the examples displayed in this paper, we used GPT version 3.5-turbo, which is the version underlying ChatGPT (and thus the version used by the most people). We have also experimented with newer OpenAI models including GPT 4 and gpt-3.5-turbo-16k. These newer models can exhibit better comprehension and analytical capabilities and handle more source passages, making them somewhat less error-prone (but also significantly more expensive to run). However, our additional experimentation indicated that the problems and biases described in this paper still persist in these newer models.

\begin{tcolorbox}[title=Box \#1: Template for prompts sent to the GPT API, float, colframe=promptbox-colframe!10!white, colback=promptbox-colframe!10!white, colbacktitle=chat-colbacklower!80!white,coltitle=black]
    \textbf{System message:} You are a helpful policy analyst working to understand the UN Biodiversity Beyond National Borders Agreement.
    \\
    \\
    \textbf{User message:} You are a helpful policy analyst working to understand the UN Biodiversity Beyond National Borders Agreement.
    \\
    \\
    \textbf{User message:} Below are some paragraphs to consider from various documents in the UN BBNJ negotiation process, including drafts of the Agreement, news bulletings about the negotiations, and submissions by various delegates:
    \\
    \\
    \begin{tcolorbox}[on line, arc=0pt]
        Relevant passages from the BBNJ document database are inserted here, having the most similar meaning to the user's question (according to the language model).
    \end{tcolorbox}
    \\
    \\
    \textbf{User message:} From information in the preceding paragraphs, please try to answer the following question. There are several drafts of the agreement leading up to the final version; please assume the question refers to the final draft unless otherwise specified.
    \\
    \\
    Question: 

    \begin{tcolorbox}[on line, arc=0pt]
        The user's question is inserted here.
    \end{tcolorbox}
    \\
    \\
    \\
    Answer:   
    
    \tcblower{
        \textbf{Box 1:} To answer the user's question using GPT, this template prompt is filled with the relevant passages found in our document database, along with the user's question, and then sent to the GPT API for completion. (The prompt is sent in a series of multiple ``user'' and ``system'' messages to help GPT distinguish the instructions it should follow from the other text in the prompt, comprising of the UN documents passages.) Box \fakeref{2} shows a filled-in example of this template.
    }
\end{tcolorbox}

\begin{tcolorbox}[float*=h, width=\textwidth, raster columns=2, title=Box \#2: Example complete prompt sent to GPT, colframe=promptbox-colframe!10!white, colback=promptbox-colframe!10!white,colbacktitle=chat-colbacklower!80!white,coltitle=black]
    \tiny
    \begin{tcolorbox}[width=(\textwidth-2mm)/2,nobeforeafter,colframe=promptbox-colframe!10!white, colback=promptbox-colframe!10!white]
    You are a helpful policy analyst working to understand the UN Biodiversity Beyond National Borders Agreement.
    \\
    \\
    Below are some paragraphs to consider from various documents in the UN BBNJ negotiation process, including drafts of the Agreement, news bulletings about the negotiations, and submissions by various delegates:
    \\
    \\
    From document "TEXTUAL PROPOSALS SUBMITTED BY DELEGATIONS BY 25 JULY 2022, FOR CONSIDERATION AT THE FIFTH SESSION OF THE INTERGOVERNMENTAL CONFERENCE ON AN INTERNATIONAL LEGALLY BINDING INSTRUMENT UNDER THE UNITED NATIONS CONVENTION ON THE LAW OF THE SEA ON THE CONSERVATION AND SUSTAINABLE USE OF MARINE BIOLOGICAL DIVERSITY OF AREAS BEYOND NATIONAL JURISDICTION (THE CONFERENCE)":
The Partiers to this Agreement,:

Rationale Added for clarity and to emphasize that this includes activities by parties or persons other than and inclusive of the project proponent

High Seas Alliance

[…] 3. Option A: “Area-based management tool” means a tool, including a marine protected area, for a geographically defined area through which one or several sectors or activities are managed with the aim of achieving particular conservation and sustainable use objectives in accordance with this Agreement. Option B: “Area-based management tool” means a tool, including a marine protected area, for a geographically defined area through which one or several sectors or activities are managed in order to achieve, in accordance with this Agreement: (a) In the case of marine protected areas, conservation objectives; (b) In the case of other area-based management tools, conservation objectives or conservation and sustainable use objectives. […] 12. “Marine protected area” means a geographically defined marine area where human activities are regulated, that is designated and managed or prohibited to achieve specific [long-term biodiversity] conservation [and sustainable use] objectives.

...

Rationale Reinserting prior text from Article 14, paragraph 4 in 2019 version as the use of common or at least similar criteria could be an important avenue for consistency, coherency and complementarity between ABMTs including MPAs adopted by States and IFBs, and for their recognition under the BBNJ Agreement.

96

High Seas Alliance

1. Areas requiring protection through the establishment of area-based management tools, including marine protected areas, shall be identified:\\
(a) On the basis of the best available science and scientific information, as well as relevant traditional knowledge of indigenous peoples and local communities, taking into account the application of precautionary principle and an ecosystem approach;

...

Rationale Measures proposed in high-seas pockets are of particular interest to those States that completely surround those measures. For one thing, a measure in a high-seas pocket might undermine conservation and sustainable use efforts taken by multiple States in the surrounding EEZs. For another, to the extent that such proposed measure will need to be established and monitored through regular cruises/trips by ships to the high-seas pocket, such cruises/trips will have to traverse the EEZ of at least one such surrounding State. Thus, for the article on consultations, it is important to recognize the particular interest of those surrounding States during the consultation process, although such States will not exercise veto power over the proposed measures. A similar concept is covered under Environmental Impact Assessments in Article 34, Option II, paragraph 6 and would be equally valid under ABMTs, including MPAs.

...

106

to the process and will allow State Parties to be constantly engaged on the developments of that MPA, permitting them to regularly consider other approaches.
We believe this proposal is balanced and could integrate the concerns of States Parties.
Article 19, Option1, paragraph 5 [placement in option I or II to be determined] 5. In cases where an area-based management tool, including a marine protected area, established under this Part subsequently falls under the sovereignty or within sovereign rights and jurisdiction the national jurisdiction of a coastal state, either wholly or in part, it shall be adapted to cover any remaining area beyond national jurisdiction or otherwise cease to be in force. It is recognised that the coastal State may decide to adopt similar measures for the maritime area concerned in its national capacity.

...

Holy See

OPTION II (keeping articles 15 and 19 separate with article 15 appearing as article 19 bis): 1. The Conference of the Parties shall take decisions on the establishment of a global network of area-based management tools, including marine protected areas, and related measures on the basis of the recommendation received from the Scientific and Technical Body regarding the final proposal and, in particular, the draft management plan, taking into account the contributions and recommendations received as a result of during the consultation process established under this Part., 1bis.The Conference of Parties shall also take decisions on such other area-based management tools, including marine protected areas that are not part of the adjoining global network in areas beyond national jurisdiction, as are appropriately proposed by a State or States for consideration and which are recommended for action by the Scientific and Technical Body after public consultation. recognizing, as appropriate, in accordance with the objectives and

...

Republic of the Philippines

Option 1

1. The Conference of the Parties shall take decisions on the establishment of area-based management tools, including marine protected areas, and related measures on the basis of the final proposal and, in particular, the draft management plan, taking into account the contributions and recommendations received during the consultation process established under this Part, recognizing, as appropriate, in accordance with the objectives and criteria laid down in this Part, area-based management tools, including marine protected areas, established under relevant legal instruments and frameworks and relevant global, regional, subregional and sectoral bodies.1. The Conference of the Parties shall take decisions on matters related to measures such as area-based management tools, including marine protected areas, with respect to proposals submitted under this Part, on a case-by-case basis and taking into account the scientific advice or recommendations and the contributions received during the consultation and assessment process.

...

\end{tcolorbox}
\begin{tcolorbox}[width=(\textwidth-2mm)/2,nobeforeafter,colframe=promptbox-colframe!10!white, colback=promptbox-colframe!10!white]

High Seas Alliance

OPTION I (merging articles 15 and 19): 1. The Conference of the Parties shall take decisions on matters related to measures such as the establishment of area-based management tools, including marine protected areas, and related measures on the basis of with respect to the final proposal and in

113

particular the draft management plans submitted under this Part, on a case-by-case basis and taking into account the scientific advice or recommendations and the contributions received during the consultation and assessment process.

[…]

5. In cases where an area-based management tool, including a marine protected area, established under this Part subsequently falls within the national jurisdiction of a coastal State, either wholly or in part, it shall be adapted to cover any remaining area beyond national jurisdiction or, at the request of the coastal State, otherwise cease to be in force.
6. A marine protected area established under this Part shall continue in force when a new regional treaty body is established with competence to establish a marine protected area that overlaps, geographically, with the marine protected area established under this Part.

...

120

High Seas Alliance

1. Parties shall ensure that activities under their jurisdiction or control that take place in areas beyond national jurisdiction are conducted and its nationals and vessels act consistently with the decisions adopted under this Part. 2. Nothing in this Agreement shall prevent a Party from adopting more stringent measures with respect to its vessels, nationals or with regard to activities under its jurisdiction or control in addition to those adopted under this Part, in conformity with international law.

[3.The implementation of the measures adopted under this Part shall not impose a disproportionate burden on small island developing States Parties, directly or indirectly.]

...

4. Following the review, the Conference of the Parties shall, as necessary, take decisions on the amendment, extension or revocation of area-based management tools, including marine protected areas, and any related measures, [as well as on the extension of timebound area-based management tools, including marine protected areas, that would otherwise automatically expire,] on the basis of the best available science and scientific information, as well as relevant traditional knowledge of indigenous peoples and local communities, taking into account the application of precaution and an ecosystem approach.

...

High Seas Alliance

1. Parties, individually or collectively, shall report to the Conference of the Parties on the implementation of area-based management tools and related measures, including marine protected areas, and related measures established under this Part. Such reports shall be made publicly available by the secretariat.

2. Area-based management tools, including marine protected areas, established under this Part, including related measures, shall be monitored by State Parties according to the monitoring and research plan, reported to the secretariat and periodically reviewed by the Scientific and Technical Body.
\\ 
\\
From document "FURTHER REVISED DRAFT TEXT OF AN AGREEMENT UNDER THE UNITED NATIONS CONVENTION ON THE LAW OF THE SEA ON THE CONSERVATION AND SUSTAINABLE USE OF MARINE BIOLOGICAL DIVERSITY OF AREAS BEYOND NATIONAL JURISDICTION":

Transparency system for benefit-sharing:

17/56 22-08354

PART III MEASURES SUCH AS AREA-BASED MANAGEMENT TOOLS, INCLUDING MARINE PROTECTED AREAS

Article 14

Identification of areas:

1. Areas requiring protection through the establishment of area-based management tools, including marine protected areas, shall be identified:

(a) On the basis of the best available science and scientific information, as

well as relevant traditional knowledge of indigenous peoples and local communities, taking into account the application of precaution and an ecosystem approach;

(b) By reference to one or more of the indicative criteria specified in annex I.

2. Indicative criteria for the identification of such areas under this Part shall include, as relevant, those specified in annex I and as may be further developed and revised as necessary by the Scientific and Technical Body for consideration and adoption by the Conference of the Parties.

Decision-making:

5. In cases where an area-based management tool, including a marine protected area, established under this Part subsequently falls within the national jurisdiction of a coastal State, either wholly or in part, it shall be adapted to cover any remaining area beyond national jurisdiction or otherwise cease to be in force.

6. A marine protected area established under this Part shall continue in force when a new regional treaty body is established with competence to establish a m arine protected area that overlaps, geographically, with the marine protected area established under this Part.
    \\
    \\
    \begin{tcolorbox}[on line, arc=0pt]
        1,248 words from 3 additional documents omitted for brevity
    \end{tcolorbox}

    \tcbline

    From information in the preceding paragraphs, please try to answer the following question. There are several drafts of the agreement leading up to the final version; please assume the question refers to the final draft unless otherwise specified.
    \\
    \\
    Question: What is the process for establishing a high seas marine protected area?
    \\
    \\
    \\
    Answer:   
\end{tcolorbox}
    
    \tcblower{
        Example prompt sent to GPT for the question in Example Chat \fakeref{1.3}: \emph{``What is the process for establishing a high seas marine protected area?''} The prompt template (Box \fakeref{1}) is filled with relevant passages from the document database.
    }
\end{tcolorbox}

%% Loading bibliography style file
\bibliographystyle{model1-num-names}
% \bibliographystyle{cas-model2-names}

% Loading bibliography database
\bibliography{bib}

%\vskip3pt

% \bio{}
% Author biography without author photo.
% Author biography. Author biography. Author biography.
% Author biography. Author biography. Author biography.
% Author biography. Author biography. Author biography.
% Author biography. Author biography. Author biography.
% Author biography. Author biography. Author biography.
% Author biography. Author biography. Author biography.
% Author biography. Author biography. Author biography.
% Author biography. Author biography. Author biography.
% Author biography. Author biography. Author biography.
% \endbio

% \bio{figs/pic1}
% Author biography with author photo.
% Author biography. Author biography. Author biography.
% Author biography. Author biography. Author biography.
% Author biography. Author biography. Author biography.
% Author biography. Author biography. Author biography.
% Author biography. Author biography. Author biography.
% Author biography. Author biography. Author biography.
% Author biography. Author biography. Author biography.
% Author biography. Author biography. Author biography.
% Author biography. Author biography. Author biography.
% \endbio

% \bio{figs/pic1}
% Author biography with author photo.
% Author biography. Author biography. Author biography.
% Author biography. Author biography. Author biography.
% Author biography. Author biography. Author biography.
% Author biography. Author biography. Author biography.
% \endbio

\end{document}